\documentclass[conference, a4paper]{IEEEtran}
\usepackage{color}
\definecolor{c}{rgb}{1,0,0}

\usepackage{cite}
\usepackage{graphicx}
\usepackage{times}
\usepackage{latexsym}

\usepackage[mathscr]{eucal}
\usepackage{array}
\usepackage{fancyhdr}
\usepackage{amsmath,amssymb}
\usepackage{bm} 
\usepackage{subfigure}
\usepackage{multirow}

\usepackage{algpseudocode}
\usepackage{algorithm}
\usepackage{amsfonts}

\ifCLASSINFOpdf
\else
\fi

\newtheorem{thm}{Theorem}

\newtheorem{prop}{Proposition}

\newtheorem{rem}{Remark}

\newcommand{\qed}{\hfill\ensuremath{\blacksquare}}

\newcommand{\nn}{\nonumber}

\newcommand{\vect}[1]{{\lowercase{\mathbf{#1}}}}
\newcommand{\mat}[1]{{\uppercase{\mathbf{#1}}}}

\newcommand{\tvec}{{\rm{vec}}}

\renewcommand{\a}{\vect{a}} % _ accent
\newcommand{\e}{\vect{e}}

\newcommand{\z}{\vect{z}}

\newcommand{\A}{\mat{A}}

\newcommand{\C}{\mat{C}}

\renewcommand{\H}{\mat{H}} % ''accent
\newcommand{\I}{\mat{I}}

\newcommand{\N}{\mat{N}}
\renewcommand{\P}{\mat{P}}

\newcommand{\R}{\mat{R}}
\renewcommand{\S}{\mat{S}}

\newcommand{\V}{\mat{V}}

\newcommand{\Z}{\mat{Z}}

\newcommand{\Rc}{{\cal R}}

\newcommand{\Ct}{{\tilde \C}}

\newcommand{\Ht}{{\tilde \H}}

\newcommand{\Pt}{{\tilde \P}}

\newcommand{\Rt}{{\tilde \R}}

\newcommand{\Cb}{{\mathbb C}}

\newcommand{\Eb}{{\mathbb E}}

\newcommand{\Lambdam}{\hbox{\boldmath$\Lambda$}}

\usepackage{xcolor}

\makeatother

\ifCLASSINFOpdf
\else
\fi

\graphicspath{{Figure/}}

\begin{document}
%
% paper title
% Titles are generally capitalized except for words such as a, an, and, as,
% at, but, by, for, in, nor, of, on, or, the, to and up, which are usually
% not capitalized unless they are the first or last word of the title.
% Linebreaks \\ can be used within to get better formatting as desired.
% Do not put math or special symbols in the title.
\title{Beam-Domain Secret Key Generation for Multi-User Massive MIMO Networks}

% author names and affiliations
% use a multiple column layout for up to three different
% affiliations
\author{\IEEEauthorblockN{You Chen\IEEEauthorrefmark{1}, Guyue Li\IEEEauthorrefmark{1}\IEEEauthorrefmark{3}, Chen Sun\IEEEauthorrefmark{2}\IEEEauthorrefmark{3}, Junqing~Zhang\IEEEauthorrefmark{4}, Eduard~Jorswieck\IEEEauthorrefmark{5}, Bin~Xiao\IEEEauthorrefmark{6}}
\IEEEauthorblockA{\IEEEauthorrefmark{1} School of Cyber Science and Engineering, Southeast University, Nanjing, 210096, China\\
\IEEEauthorrefmark{2}National Mobile Communications Research Laboratory, Southeast University, Nanjing, 210096, China\\
\IEEEauthorrefmark{3}Purple Mountain Laboratories for Network and Communication Security, Nanjing, 210096, China\\
\IEEEauthorrefmark{4}Department of Electrical Engineering and Electronics, University of Liverpool, Liverpool, L69 3GJ, United Kingdom\\
\IEEEauthorrefmark{5}Institute for Communications Technology, Technische Universit\"at Braunschweig,  Germany\\
\IEEEauthorrefmark{6}Department of Computing, The Hong Kong Polytechnic University, Hong Kong\\
Corresponding author: Guyue Li, Email: \{guyuelee\}@seu.edu.cn
}
}

% conference papers do not typically use \thanks and this command
% is locked out in conference mode. If really needed, such as for
% the acknowledgment of grants, issue a \IEEEoverridecommandlockouts
% after \documentclass

% for over three affiliations, or if they all won't fit within the width
% of the page, use this alternative format:
%
%\author{\IEEEauthorblockN{Michael Shell\IEEEauthorrefmark{1},
%Homer Simpson\IEEEauthorrefmark{2},
%James Kirk\IEEEauthorrefmark{3},
%Montgomery Scott\IEEEauthorrefmark{3} and
%Eldon Tyrell\IEEEauthorrefmark{4}}
%\IEEEauthorblockA{\IEEEauthorrefmark{1}School of Electrical and Computer Engineering\\
%Georgia Institute of Technology,
%Atlanta, Georgia 30332--0250\\ Email: see http://www.michaelshell.org/contact.html}
%\IEEEauthorblockA{\IEEEauthorrefmark{2}Twentieth Century Fox, Springfield, USA\\
%Email: homer@thesimpsons.com}
%\IEEEauthorblockA{\IEEEauthorrefmark{3}Starfleet Academy, San Francisco, California 96678-2391\\
%Telephone: (800) 555--1212, Fax: (888) 555--1212}
%\IEEEauthorblockA{\IEEEauthorrefmark{4}Tyrell Inc., 123 Replicant Street, Los Angeles, California 90210--4321}}

% use for special paper notices
%\IEEEspecialpapernotice{(Invited Paper)}

% make the title area
\maketitle

% As a general rule, do not put math, special symbols or citations
% in the abstract
\begin{abstract}
Physical-layer key generation (PKG) in multi-user massive MIMO networks faces great challenges due to the large length of pilots and the high dimension of channel matrix. To tackle these problems, we propose a novel massive MIMO key generation scheme with pilot reuse based on the beam domain channel model and derive close-form expression of secret key rate. Specifically, we present two algorithms, i.e., beam-domain based channel probing (BCP) algorithm and interference neutralization based multi-user beam allocation (IMBA) algorithm for the purpose of channel dimension reduction and multi-user pilot reuse, respectively. Numerical results verify that the proposed PKG scheme can achieve the secret key rate that approximates the perfect case, and significantly reduce the dimension of the channel estimation and pilot overhead.
\end{abstract}

% no keywords

% For peer review papers, you can put extra information on the cover
% page as needed:
% \ifCLASSOPTIONpeerreview
% \begin{center} \bfseries EDICS Category: 3-BBND \end{center}
% \fi
%
% For peerreview papers, this IEEEtran command inserts a page break and
% creates the second title. It will be ignored for other modes.
\IEEEpeerreviewmaketitle

\section{Introduction}
% no \IEEEPARstart
\label{sec:intro}
%Nowadays, with the rapid development of fifth generation (5G) and beyond communication systems, their demand for supporting extremely high throughput and multiple users makes massive MIMO technology a must in the future~\cite{8241348}.

The fifth generation (5G) and beyond communication systems have been developing at an unprecedented speed to meet the requirement of high data rate and low latency.
In the 5G networks, the random access from tons of devices makes traditional cryptographic key distribution and management very challenging. Under this background, the physical-layer key generation (PKG) has emerged as an alternative technique to establish the symmetric key for cryptographic applications~\cite{Li2019Physical}.
PKG can generate time-varying key with a lightweight algorithm from the channel randomness.
Thanks to the channel decorrelation property, there is no information leakage to eavesdroppers, when they are located half a wavelength away or more from legitimate users{~\cite{Jorswieck2014Secret}}.

%Most of existing PKG schemes have been focused on the single antenna or small-scale multiple-input multiple-output (MIMO) communication systems~\cite{Jorswieck2014Secret}.

%However, intuitively extending existing pairwise PKG approach in small scale antennas scenarios to the multi-user massive MIMO scenarios will face great challenges.
%However, existing pairwise PKG approach in small scale antennas scenarios cannot be directly applied in massive MIMO network due to two main reasons as follows. Firstly, it brings in large length of pilots to distinguish different users, while it is hard to accomplish channel probing within the coherence time in a time division duplex (TDD) system. Secondly, the high dimension of channel matrix makes it with high computational complexity to perform the decorrelation preprocessing algorithms such as principal component analysis~\cite{Li2018High}.

The 5G networks employ massive MIMO technology to support extremely high throughput and multi-user access.  However, traditional pairwise PKG method is difficult to scale to the new scenario due to the high dimension of channel matrix caused by massive MIMO antennas and the huge number of  orthogonal pilot overhead to distinguish multiple users~\cite{Li2019Physical}.
%Few work has investigated PKG in multi-user massive MIMO networks.
Jiao~\textit{et al.}  proposed   to use new channel characteristics, i.e., virtual angle of arrival (AoA) and angle of departure (AoD), to generate a shared secret key for pairwise users in a massive MIMO system~\cite{Jiao2}. However, they only considered the PKG between two legitimate users. %It is still missing how to generate secret key with multiple users.
Generating secret keys between a base station and multiple users has been a largely under explored domain~\cite{zhang2016review}. Several work studied  the group PKG protocols, where all users in the group negotiate a common key based on their channel estimates.  But the majority of them still perform channel probing in a pairwise manner, resulting in an extremely large overhead and low efficiency.
Hence, those works related to PKG among multiple nodes through the optimization of probing rates at individual node pair and channel probing schedule do not scale in this context.
Exceptionally, Zhang~\textit{et al.} designed a multi-user key generation protocol by leveraging the multi-user access of OFDMA modulation, which is achieved by assigning non-overlapping subcarriers to different users~\cite{zhang2019design}. However, no work has been devoted to exploiting the spatial diversity of massive MIMO to enable multi-user key generation.

In summary, it is still missing how to generate secret keys among multiple users in massive MIMO networks, which is tackled in this paper.
The main contributions are as follows:
\begin {itemize}
\item  We propose a channel dimension reduction approach by exploiting the spatial sparse property of the beam domain channel model.
Employing this approach, legitimate users only need to estimate the effective channels at a few dominant beams, which allows us to reduce the dimension of channel estimation significantly.
\item
% We derive a closed-form expression of the secret key rate, considering other UTs as non-colluding curious users.
%Furthermore, w
We propose a muti-user pilot reuse approach that can largely reduce the pilot overhead compared with using orthogonal signals.
Furthermore, we present a novel algorithm to design the precoding and receiving matrices, with the purpose of achieving the perfect secret key rate.
\item Numerical results verify that the approach can
achieve the secret key rate that approximates the perfect case and significantly reduce the dimension of the channel estimation
and the pilot overhead.
 \end {itemize}
\section{Secret key generation with MU massive MIMO} % (fold)
\label{sec:MUMASSIVE}
\subsection{System Model and Problem Statement}
\begin{figure}[!t]
\centering
{\includegraphics[width=0.35\textwidth]{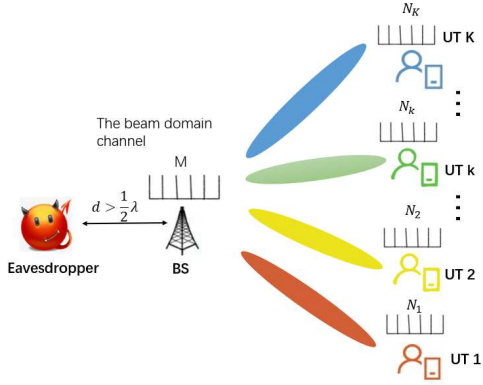}}
\caption{System model of multi-user secret key generation.}
\label{fig:system}
\end{figure}

This paper considers a narrow-band star topology network, where a base station (BS) simultaneously generates secret keys $\kappa = \{\kappa_1, \kappa_2, \cdots, \kappa_K\}$ with $K$ user terminals (UT), as shown in Fig.~\ref{fig:system}.
The BS is equipped with $M$ antennas and the $k$-th UT is equipped with $N_{k}$ antennas. We consider the potential unintended hearing from other UTs and the active attacks are out of scope in this paper.

%The fifth generation (5G) and beyond communication systems employ massive MIMO technologies to support extremely high throughput and multiple users~\cite{8241348}.
%However, it is challenging to apply PKG with massive MIMO~\cite{Li2019Physical}.

%Most of existing PKG schemes have been focused on the single antenna or small-scale multiple-input multiple-output (MIMO) communication systems~\cite{Jorswieck2014Secret}.
%With the rapid development of fifth generation (5G) and beyond communication systems, their demand for supporting extremely high throughput and multiple users makes MU massive MIMO technology a must in the future~\cite{8241348}. However, intuitively
It is challenging for existing pairwise PKG approach to scale to multi-user massive MIMO scenarios due to two main reasons as follows.
\begin{enumerate}
\item {\textit{High dimension of the channel matrix:}} The elements of the generated secret keys are highly auto-correlated due to the spatial correlation of the antennas, which must be reduced by decorrelation preprocessing algorithms.
However, in a multi-user massive MIMO network, the high dimension of channel matrix makes it too complicated to perform the decorrelation preprocessing algorithms such as principal component analysis.
   % , {\color{red}resulting in high auto-correlated elements of the generated secret keys.}

%can be too high to Because of the spatial correlation of the antennas, the elements of the generated secret keys are highly are highly auto-correlated. However, due to the large scale of antennas at  the BS and $\z_k^{UL}$ and $\z_k^{DL}$ with a large dimension of $M N_k$, the traditionally auto-correlation reducing algorithm, e.g., PCA, may be too complicated to perform.
%     the high
%dimension of channel matrix makes it with high computa-tional complexity to perform the decorrelation preprocessing
%algorithms such as principal component analysis [4].

\item {\textit{Large pilot overhead:}} The length of uplink pilots scales with the number of antennas as well as the number of UTs. Thus, in the massive MIMO network where the number of antennas is extremely large, it brings in large length of pilots to distinguish different users, while it is hard to accomplish channel probing within the coherence time in a time division duplex (TDD) system.

   % is impractical for the BS and the user terminal (UT) to estimate the instantaneous uplink and downlink channel information within the channel coherence time. This indicates that the BS and UT may not obtain highly correlated CSI

%    .%the since the base station (BS) is equipped with extremely large number of antennas in massive MIMO systems,

% the length of uplink pilots scales with the number of antennas $N_k$ as well as the number of UTs $K$.
%   Therefore, the minimum time of channel probing is $T= (M+N) \Delta T +T_{Switch}$, where $\Delta T$ is the symbol transmission time, and $T_{Switch}$ is the switching time from downlink to unlink.
%   This time needs to be deliberately kept smaller than the channel coherence time, so that BS and UTs can obtain highly correlated CSI in a TDD system.
%   When $M$, $K$ and $N_k$ are large, it becomes very challenge to accomplish channel probing within the coherence time.
\end{enumerate}
\subsection{Scheme Framework}
\label{sub:secret_key_rate}
The proposed framework for multi-user PKG is portrayed in Fig.~\ref{fig:mmW}.
It contains four steps, namely channel probing, quantization, information reconciliation, and privacy amplification. The last three steps are similar with existing work summarized in~\cite{Li2019Physical}, so this paper will focus on the first step, i.e., channel probing, which is relatively different from that in the point-to-point PKG.
 {To address the two challenges mentioned  above, the proposed channel probing scheme contains two novel parts, i.e.,  channel dimension reduction and  multi-user pilot reuse.
 %beam-domain transform based channel dimension reduction (BCDR) algorithm and  interference neutralization based multi-user beam allocation (IMBA)  algorithm.}
%Also note that we consider single antenna at UT side, so the design of the receiving matrix of UT is beyond the  scope of this paper.
\begin{figure}[!t]
\centering
\includegraphics[width=0.4\textwidth]{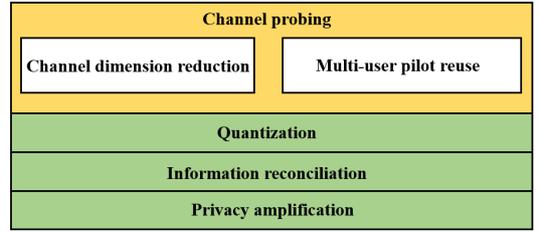}\label{fig:1}
%\subfigure[Uplink]{\includegraphics[width=0.45\textwidth]{2.pdf}\label{fig:2}}
\caption{Framework of secret key generation scheme in the multi-user massive MIMO network.}\label{fig:mmW}
\end{figure}

%We consider the following two steps in the key generation scheme:

%\begin{enumerate}
%   \item {\textit{Beam allocation:}} In this step, BS and UTs design the precoding matrix according to their statistical CSI.
%  Firstly, in the uplink, each UT transmit the sounding signals. Then, the BS estimates the covariance matrix and designs the precoding matrix $\P_k$. {\color{blue} By employing the precoding matrix $\P_k$, BS can select the strongest beams in the downlink transmission.}
%  %Next, in the downlink, the BS employs the precoding matrix to transmit the downlink sounding signals.Then, each UT estimates the covariance matrix and designs the receiving matrix $\C_k$.
%    \item {\textit{Channel probing:}} In this step, BS and UTs probe the channel alternatively and construct the reciprocal channel characteristics using the precoding matrix.
%  Firstly, the BS transmits the downlink pilot signals by the precoding matrix $\P$. Then UTs receive the received signals
% % by the matrix $\C^H$
%  and obtain the reciprocal channel parameters. Next, each UT %employs the matrix $\C^*$
%  transmit the pilot signals to BS. The BS utilizes the precoding matrix $\P^{T}$ to preprocess the received signals and estimate the effective channel parameters.
%
%\end{enumerate}
{\begin{enumerate}
  \item {\textit{Channel dimension reduction:}}
      In the massive MIMO channel matrix, only a few dominant elements contain the most relevant channel information.
      So we employ the beam domain channel model where the channel gains are concentrated in a few beams. In order to significantly reduce the dimension of the channel estimation, a  beam-domain based channel probing (BCP) algorithm is proposed to obtain the CSI. This scheme will be presented in Section III.
  \item {\textit{Multi-user pilot reuse:}} In order to reduce the pilot overhead in massive MIMO network, we consider pilot reuse among the UTs, where different UTs transmit the identical pilot signals. With the purpose of mitigating the inter-user interference, we then present an interference neutralization based beam allocation algorithm (IMBA) to then proposed to design the precoding and receiving matrix. This algorithm will be discussed further in Section IV.
\end{enumerate}}

\section{{Channel dimension reduction approach }} % (fold)
\subsection{{Beam-domain Channel Model}} % (fold)

%%%%%%%%%%%%%%%%%%%%%%%%%%%%%%%%%%%%%%%%%%%%%%%%%%%%
%\begin{align}\label{eq:reciprocal}
%    \H_{k}^{UL}= (\H_{k}^{DL})^T,
%\end{align}
%%%%%%%%%%%%%%%%%%%%%%%%%%%%%%%%%%%%%%%%%%%%%%%%%%%%%
We consider a narrow-band multipath channel model.
The downlink channel response of the $k$-th UT
can be given as
\begin{align}\label{eq:1}
    \H_{k}^{DL}= \sum_{p=1}^{N_P} \H_{k,p}^{DL}
    &=\sum_{p=1}^{N_P}\alpha_{k,p} \a_{UT,k}(\theta_{k,p}) \a^H_{BS}(\varphi_{k,p}).
\end{align}
where $N_P$ is the number of paths, $\H_{k,p}^{DL}$ is the
%$N_k\times M$
 downlink channel matrix associated with the $p$-th path of $k$-th UT~\cite{DTse}, $\alpha_{k,p}$ is the complex gain of the $p$-th path,
$\a_{UT,k}(\theta_{k,p})$ and $\a_{BS}(\varphi_{k,p})$ are the antenna array response vectors at the UT and BS with AoA $\theta_{k,p}$ and AoD $\varphi_{k,p}$, respectively.
Specifically, under the uniform linear array (ULA) setup,
these vectors are given by
\begin{align}
    \a_{UT,k}(\theta_{k,p}) =\frac{1}{\sqrt{N_{k}}} \left[
    1, e^{-j\Theta_{k,p}},\ldots, e^{-j(N_k-1)\Theta_{k,p}}\right]^T\nn\\
    \a_{BS}(\varphi_{k,p})=\frac{1}{\sqrt{M}} \left[
    1, e^{-j\Phi_{k,p}},\ldots,e^{-j(M-1)\Phi_{k,p}}\right]^T,
\end{align}
where $\Theta_{k,p}=\frac{2\pi}{\lambda}d \sin(\theta_{k,p})$, $\Phi_{k,p} = \frac{2\pi}{\lambda}d \sin(\varphi_{k,p})$, $\lambda$ is the wavelength, and $d$ is the distance between the adjacent antennas.

Beam domain model samples the original physical channel by two series of uniformly distributed beams/angles over $[0,2\pi]$, i.e., transmitting and receiving beams/angles.
According to \cite{BDMA}, the downlink beam domain channel response is
\begin{align}\label{eq:6}
    \Ht_{k}^{DL} = \A_{UT,k}^H \H_{k}^{DL} \A_{BS},
\end{align}
where
$\A_{UT,k} = \left[\a_{UT,k}(\theta_1),\a_{UT,k}(\theta_2),\ldots,\a_{UT,k}(\theta_{N_k})  \right]  \in \Cb^{N_k\times N_k}$
and
$\A_{BS} = \left[\a_{BS}(\varphi_1),\a_{BS}(\varphi_2),\ldots,\a_{BS}(\varphi_M)  \right]  \in \Cb^{M\times M}$
% \begin{align}
%     \A_{UT,k} = \left[\a_{UT,k}(\theta_1),\a_{UT,k}(\theta_2),\ldots,\a_{UT,k}(\theta_{N_k})  \right]  \in \Cb^{N_k\times N_k}
% \end{align}
% and
% \begin{align}
%     \A_{BS} = \left[\a_{BS}(\varphi_1),\a_{BS}(\varphi_2),\ldots,\a_{BS}(\varphi_M)  \right]  \in \Cb^{M\times M}
% \end{align}
are the sampling matrices at the $k$-th UT and the BS, respectively.
%They satisfy that $ \A_{UT,k}^H \A_{UT,k} = \I$, $\A_{BS}^H \A_{BS}=\I $.
The $(n,m)$-th element of $\Ht_{k}^{DL}$ represents the channel gains from AoD $\varphi_m$ to  AoA $\theta_n$, where $\varphi_m$ and $\theta_n$ are the $m$-th and $n$-th sample angles, which satisfy that $\sin(\varphi_m) = 2m/M-1$ and $\sin(\theta_n) = 2n/N_k-1$.
%When the antenna spacing is half wavelength, i.e., $d = \lambda/2$,
%the matrices $\A_{UT}$ and $\A_{BS}$ become the unitary discrete Fourier transform (DFT)
%matrix, defined as \cite{7913686}
%\begin{align}
%    [\A_{UT,k}]_{n_1,n_2} = \frac{1}{\sqrt{N_k}} \exp(-j2\pi(n_1-1)(n_2-N_k/2)/N_k)\nn\\
%    [\A_{BS}]_{m_1,m_2} = \frac{1}{\sqrt{M}} \exp(-j2\pi(m_1-1)(m_2-M/2)/M).
%\end{align}

%When the number of antennas tends to infinity,
%the beam domain channel $\Ht_{k}^{DL}$ exhibits some interesting and favorable properties.
\begin{prop}\label{prop:1}
    When the number of antennas grows to infinity,
    the $(n,m)$-th element of beam domain channel $\Ht_{k}^{DL}$ tends to~\cite{BDMA}
    \begin{align}
    \lim_{M,N_k\to \infty}&\Big( [\Ht_{k}^{DL} ]_{n,m} -\sum_{p=1}^{N_P} \alpha_{k,p} \delta(\theta_{k,p} - \arcsin(2n/N_k-1))\nn\\
       & \times\delta(\varphi_{k,p} - \arcsin(2m/M-1))\Big) = 0.
    \end{align}
The beam domain channel covariance matrices
$\Rt_{BS,k}= \Eb \{ (\Ht_{k}^{DL})^H \Ht_{k}^{DL} \}$ and $\Rt_{UT,k}=\Eb \{  \Ht_{k}^{DL} (\Ht_{k}^{DL})^H \}$ tend to diagonal matrices with the diagonal elements given by
\begin{small}
\begin{align}\label{eq:9}
    \lim_{M\to\infty} [\Rt_{BS,k}]_{m,m}- \sum_{p=1}^{N_P} |\alpha_{k,p}|^2
        \delta(\varphi_{k,p} - \arcsin(2m/M-1)) \nn = 0,\nn\\
    \lim_{N_k\to\infty} [\Rt_{UT,k}]_{n,n} -\sum_{p=1}^{N_P} |\alpha_{k,p}|^2 \delta(\theta_{k,p} - \arcsin(2n/N_k-1))\nn = 0.\\
\end{align}
\end{small}
\end{prop}

\begin{rem}
%Proposition~\ref{prop:1} reveals that when the BS and UT $k$ are equipped with a large number of antennas, the beam domain channel matrix $\Ht_{k}^{DL}$ tends to matrix $\G_{k}^{DL}$. In practice, when the BS and UT $k$ are equipped with a large (but finite) number of antennas, the beam domain channel matrix $\Ht_{k,p}^{DL}$ can be approximated by $\G_{k,p}^{DL}$.
% From the definition of $\G_{k}^{DL}$ in \eqref{eq:G},
%for each $n$ and $m$, there is at most one path $p$ simultaneously satisfying $\theta_{k,p} = \arcsin(2n/N_k-1)$ and $\varphi_{k,p} = \arcsin(2m/M-1)$.
%This indicates that one element in $\G_{k}^{DL}$ represents channel gains from one AoD $\varphi_{k,p}$ to one AoA $\theta_{k,p}$ and
%different elements represent channel gains corresponding to different AoAs and AoDs.
%As there are $N_P$ paths, the number of non-zero entries in matrix $\G_k^{DL}$ is $N_P$.
For each $n$ and $m$, there is at most one path $p$ simultaneously satisfying $\theta_{k,p} = \arcsin(2n/N_k-1)$ and $\varphi_{k,p} = \arcsin(2m/M-1)$, which means that
different elements represent channel gains corresponding to different AoAs and AoDs.
With a large (but finite) number of antennas, $\Ht_k^{DL}$ is a very sparse matrix with $N_P$ dominant elements corresponding to the paths.
Moreover, these elements become independent with each other as long as these paths are independent.
The $m$-th diagonal element in $\Rt_{BS,k}$ represents the channel gains of the $m$-th transmit beam ($\varphi_{k,p} = \arcsin(2m/M-1)$), and the $n$-th diagonal element in $\Rt_{UT,k}$ represents the channel gains of the $n$-th receive beam ($\theta_{k,p} = \arcsin(2n/N-1)$).
\end{rem}

\subsection{{The BCP Algorithm}} % (fold)
\label{sec:BCP}
%\begin{enumerate}
%  \item {\textit{}} In this step, BS and UTs design the precoding and receiving matrix according to their statistical CSI.
%Firstly, in the uplink, each UT employs one antenna to transmit the sounding signals. Then, the BS estimates the covariance matrix and designs the precoding matrix $\P$.
%Next, in the downlink, the BS employs the precoding matrix to transmit the downlink sounding signals and UTs estimate the statistical CSI information the covariance matrix at the UT side and design the receiving matrix $\C$.

    %\item {\textit{}}
    {In this section, we assume the precoding and receiving matrices have been provided, the algorithm of designing these matrices will be presented later in Section V.

    In this stage, BS and UTs probe the channel alternatively and employ the precoding and receiving matrix to construct the reciprocal channel characteristics. Firstly, the BS transmits the downlink pilot signals by the precoding matrix $\P$ and UTs preprocess the received signals by the receiving matrix $\C^H$ to obtain the reciprocal channel parameters. Next, each UT employs the matrix $\C^*$ to transmit the pilot signals. The BS utilizes the matrix $\P^T$ to preprocess the received signals and estimate the effective channel.}

    Based on the analysis above, we propose a BCP algorithm, which is illustrated in Algorithm \ref{alg:1}.

%\end{enumerate}

\subsection{Signal Presentation} % (fold)
\label{sec:CDR-KGS}

% The downlink and uplink probing process is illustrated in Fig.~\ref{fig:mmW}.
% Let $\P= [\P_1,\P_2,\cdots,\P_K]$ and $\C^H= [\C_1^H,\C_2^H,\cdots,\C_K^H]$ denote the precoding and receiving matrices in the downlink transmission, respectively.
% Then, $\C^*$ and $\P^T$ are used as precoding and receiving matrices in the uplink transmission.
% The dimensions of $\P_k$ and $\C_k^H$ are $M\times M_e$ and $N_e\times N_k$, respectively, where $N_e$ and $M_e$ represent the dimension of the effective channel matrix at the UT and BS sides. In general, $N_e$ and $M_e$ are approximately equal to the number of paths $N_P$, which is far smaller than $M$ and $N_k$.
% Thus, it can effectively reduce the dimension.

In the downlink transmission, define the downlink pilot from BS to UT $k$ within $T_D$ time slots as $\S_k^{DL}\in \Cb^{M_e\times T_D}$, where $M_e$ is the dimension of the effective channel at the BS. To estimate the perfect CSI, the pilot signals of each UT are orthogonal.
%, which are reused between UTs, given by $\S_{k'}^{DL}(\S_k^{DL})^H = \I$.
%After multiplying the pilot signals by the precoding matrix $\P_k\in\Cb^{M\times M_e}$, the BS transmits the summation of all the {\color{red}signals}. Then, UT $k$ multiplies received signal by the receiving matrix $\C_k^H\in\Cb^{N_e\times N_k}$, where $N_e$ is the dimension of the effective channel at the UT.
%By the LS estimation, UT $k$ estimates the downlink CSI as
{Based on the BCP algorithm, the downlink CSI estimated at  UT $k$ side is
\begin{align}\label{eq:21}
   \!\!\!\! \Z_k^{DL} =
    %\Y_k^{DL} (\S_k^{DL})^H =
 \C_k^H \H_k^{DL}\! \sum_{k'}\P_{k'}\S_{k'}^{DL} (\S_k^{DL})^H
+ \C_k^H\N_k (\S_k^{DL})^H.
\end{align}

In the uplink transmission, define the pilot transmitted by UT $k$ within $T_U$ time slot as $\S_k^{UL}\in\Cb^{N_e\times T_U}$, which satisfies $\S_{k'}^{UL} (\S^{UL}_k)^H = \I$.
%The $k$th UT employs the matrix $\C_k^*$ to transmits pilot signals, and the BS receives the summation of all the UTs' signals.
%After multiplying by the receiving matrix $\P_k^T$
%and employing the LS estimation, the estimated effective channel of UT $k$ can be expressed as
Employing the BCP algorithm, the estimated uplink CSI of UT $k$ can be expressed as
\begin{align}\label{eq:19}
    \Z_k^{UL} =
    %\Y^{UL} (\S^{UL}_k)^H =
   \P_k^T\sum_{k'} \H_{k'}^{UL}\C_{k'}^* \S_{k'}^{UL} (\S^{UL}_k)^H
     + \P_k^T\N (\S^{UL}_k)^H.
\end{align}
}
\begin{algorithm}[h]
\caption{BCP algorithm.}
\label{alg:1}
\begin{algorithmic}[1]
\Require $\P_{k}$ and $\C_{k}$
\Ensure $\z_k^{UL}$ and $\z_k^{DL}$
\State \textbf{In the downlink:}
\State \textbf{At the BS side:}
\For{$k=1:K$}
\State  Multiply the downlink pilot signals $\S_k^{DL}$ to UT $k$ by the precoding matrix $\P_k$ .
\EndFor
\State  Transmit the summation of all processed signals to UTs.

\State \textbf{At the UT side:}
\For{$k=1:K$}
\State  UT $k$ multiplies received signal by the receiving matrix $\C_{k}^H$ and employ the LS estimation to estimate the downlink CSI $\Z_k^{DL}$.
\State Vectorize the estimated effective channel matrices $\Z_k^{DL}$ as $\z_k^{DL} = \tvec(\Z_k^{DL})$
\EndFor
\State \textbf{In the uplink:}
\State \textbf{At the UT side:}
\For{$k=1:K$}
\State  Multiply the uplink pilot signals $\S_k^{UL}$ to the BS by the matrix $\C_k^*$.
\State Transmit the summation of all processed signals to the BS.
\EndFor
\State \textbf{At the BS side:}
\For{$k=1:K$}
\State  Multiply received signal by the matrix $\P_{k}^T$ and employ the LS estimation to estimate the uplink CSI $\Z_k^{UL}$.
\State Vectorize the estimated effective channel matrices $\Z_k^{UL}$ as $\z_k^{UL} = \tvec(\Z_k^{UL})$
\EndFor
\end{algorithmic}
\end{algorithm}
% \begin{multline}\label{eq:19}
%     \Z_k^{UL} =
%     %\Y^{UL} (\S^{UL}_k)^H =
%     \P_k^T\H_{k}^{UL}\C_{k}^*
%     +\sum_{k'\ne k} \H_{k'}^{UL}\C_{k'}^*\S_{k'}^{UL} (\S^{UL}_k)^H
%     \\ + \P_k^T\N (\S^{UL}_k)^H.
% \end{multline}
\begin{rem}
In the uplink and downlink transmissions, the BS and UTs vectorize the estimated effective channel matrices as $\z_k^{DL} = \tvec(\Z_k^{DL})$ and $\z_k^{UL} = \tvec((\Z_k^{UL})^T)$  to generate the secret key.
As the uplink and downlink channels are reciprocal, the downlink channel $\H_k^{DL}$ is denoted as $\H_k$, and the uplink channel is $\H_k^{UL}=(\H_k)^T$.
The reciprocal component between the BS and UT $k$ is $\C_k^H \H_k \P_k$ with a small dimension of $N_e \times M_e$.
In this way, the dimension of channel characteristics is reduced by $\eta = \frac{M\times N_k}{N_e \times M_e}$ times. The dimensions of $M_e$ and $N_e$ are very small compared with the number of antennas, therefore the dimension can be significantly reduced.
\end{rem}

\subsection{Secret Key Rate}
\label{sub:secret_key_rate}

%
%When the BS communicates with one UT, other UTs are potential non-colluding curious users\footnote{The case of colluding eavesdroppers could be included as well. Due to space constraints, we focus on non-colluding scenario here.}.
%Under the TDD operation, each UT cannot transmit and receive signals at the same time. The $i$th UT only has the channel observation in the downlink transmission.
%Thus, the key rate is the minimum mutual information given other UT's observations. The number of secure bits for the link from the BS to UT $k$ in the mutual information can be expressed as \cite{Liang2009}
%\begin{align}\label{eq:27}
%    I_{k} = \min_{i\neq k} I (\z_k^{DL};\z_k^{UL} | \z_i^{DL}  ).
%\end{align}

This paper considers the condition that the beam domain channel between one UT and the BS is independent of that between one UT and another UT. Thus, the secret key rate is the minimum mutual information between $\z_k^{DL}$ and $\z_k^{UL}$, which can be expressed as $I_k = I (\z_k^{DL};\z_k^{UL} )$.
%In massive MIMO communications, when the beam domain channels of different UTs are non-overlapping, i.e., the channel covariance matrices at the BS are orthogonal, given by
%\begin{align}
%     \Rt_{BS,k}\Rt_{BS,i} = {\bf 0},\quad k\neq i,
% \end{align}
%the channel vectors of UT $k$ and UT $i$ are independent. Then, the secret rate $I(\z_k^{DL};\z_k^{UL}|\z_i^{DL})$ can be degraded as $I(\z_k^{DL};\z_k^{UL})$ and no secret keys are leaked to potential curious UTs~\cite{6584929}.
%
%When the beam domain channels are overlapping, we should consider the information leakage to other UTs. But we can always select non-overlapping beams for different UTs and then the selected channel information is independent. Thus, we can also use (\ref{eq:degrade}) to calculate the secret key rate. The overlapping case will be discussed in more detail in Section~\ref{sub:security_analysis}.
%
%
%
%% Noting that in this paper, we only consider the orthogonal pilot for each UT and the pilot signals between UTs are not orthogonal. The length of orthogonal pilot for each UT depends on the dimensions of the precoding and receiving matrices. Thus, it is reasonable to assume orthogonal pilot by decrease the dimensions. As the number of UTs increase, if we use orthogonal pilots between UTs, the pilot length scales with the number of UTs. Hence, we do not restrict the orthogonal pilot between UTs.
Denote the precoding and receiving matrices in the beam domain as $\Pt_k = \A_{BS}^H\P_{k}$ and $\Ct_k = \A_{UT,k}^H\C_k$, respectively.
Let $\V_k = \Lambdam_k^{1/2}\Big(\sum_{k'}(\Pt_{k'})^T \otimes \Ct_k^H\Big)^H$ and $\V_{kk'} = \Lambdam_{k'}^{1/2} \Big(\Pt^T_k\otimes \Ct^H_{k'}\Big)^H$,
where $\Lambdam_k = \Eb\{\tvec(\Ht_k)\tvec(\Ht_k)^H\}$ is the full correlation of the beam domain channel.
\begin{thm}\label{thm:1}
When the channel of different UTs are independent, according to~\cite{BDMA}, we can compute the secret key rate of UT $k$ as
    \begin{align}\label{eq:53}
        &I_k = \!- \log\! \det\! \Bigg(\!\I - \V_{kk}\!  \left(\! \sum_{k'} \V_{kk'}^H \V_{kk'}  \!+\! \Big(\P^T_k\P^*_k \otimes \I_{T_U}\!\Big)\!\!\right)^{-1}\nn\\
    &\quad \times \V_{kk}^H \V_k \left(  \V_k^H\V_k + \I_{T_D} \otimes \C_k^H \C_k \right)^{-1}\V_k^H \Bigg).
    \end{align}
\end{thm}
\begin{IEEEproof}
    See Appendix \ref{sec:proof_of_theorem_thm:1}.
\end{IEEEproof}

\section{Multi-user pilot reuse approach}

%In this section, we consider the CDR-based secret key generation scheme design under the pilot reuse case, where different UTs transmit the identical pilot signals.
%In this case, we first design the precoding and receiving matrices maximizing the secret key rate and then analyze the security when the channels of different UTs are correlated.

% In this case,  the secret key rate can be expressed as
%     \begin{align}\label{eq:37}
%         &I(\z_k^{DL};\z_k^{UL}) =  -\log \det \left(  \I - \Lambdam_k^{1/2} \Big(\Pt^T_k\otimes \Ct^H_{k}\Big) \right.  ^H\nn\\
%         &   \times \left(\I + \sum_{k'}\Big(\Pt^T_k\otimes \Ct^H_{k'}\Big)
%      \Lambdam_{k'} \Big(\Pt^T_k \otimes \Ct^H_{k'}\Big) ^H\right)^{-1} \nn\\
%     &  \times \Big(\Pt^T_k\otimes \Ct^H_{k}\Big)
%     \Lambdam_k\Big(\sum_{k'}(\Pt_{k'})^* \otimes \Ct_k\Big)\nn\\
%     &\times  \left( \I + \Big(\sum_{k'}(\Pt_{k'})^T \otimes \Ct_k^H\Big) \Lambdam_k\Big(\sum_{k'}(\Pt_{k'})^* \otimes \Ct_k\Big)  \right)^{-1}\nn\\
%     &\left.\times\Big(\sum_{k'}(\Pt_{k'})^T \otimes \Ct_k^H\Big) \Lambdam_k^{1/2}\right).
%     \end{align}

% \subsection{Approach Description}

% From the above analysis, we find that 1) it is optimal to generate secret keys in the beam domain; 2) the beams of different UTs should be non-overlapping.

\subsection{Pilot Reuse for Multi-users} % (fold)
\label{sec:CDR-KGS}
According to \eqref{eq:53}, when the pilot signals of each UT are orthogonal, there is no inter-user interference and the estimated CSI achieves the perfect case.
The pilot overhead is defined as the length of the total  pilot signals.
For traditional approach using the orthogonal pilot, the pilot overhead is given by
\begin{align}
	T_{TA}= M + \sum\limits_{k = 1}^K {{N_k}}.
\end{align}
However, as $T_{TA}$ scales with the number of antennas and users, the overhead of orthogonal signals is extremely large in the multi-user massive MIMO network.
Therefore, we consider the key generation scheme under the pilot reuse case, where different UTs transmit the identical pilot signals. In the pilot reuse case, the pilot overhead is reduced to
\begin{align}
T_{PA}= {M_e} + {N_e}.	
\end{align}

Since pilot reuse results in interference between UTs and reduces secret key rate, it is necessary to mitigate the interference.

{
The interference neutralization approach can be employed to reduce the interference, i.e., for arbitrary matrix $\Ct_{k'}$ ($k'\neq k$), the precoding matrix $\Pt_k$ satisfies
\begin{align}\label{eq:48}
    (\Pt_k^T\otimes \Ct_{k'}^H) \Lambdam_{k'} = {\bf 0},\quad k'\neq k.
\end{align}

When the channel beams of different users are non-overlapping, we have
\begin{align}\label{eq:49}
    \Pt_k^H \Rt_{BS,k'}= {\bf 0},\quad k'\neq k.
\end{align}}The constraint of interference neutralization approach can always be satisfied under this case.
%the constraint \eqref{eq:48} can be easily satisfied.
%Under this constraint, problem \eqref{eq:p0} maximizing the sum secret key rate can be decomposed into the following subproblems maximizing the secret key rate of each user
%\begin{align}\label{eq:prob1}
%    \min_{\Pt_k,\Ct_k} \  & \log \det \Big( \I - \Lambdam_k^{1/2} \Big(\Pt^* _k\otimes \Ct_{k}\Big)  \Big(\I + \Big(\Pt^T_k\otimes \Ct^H_{k}\Big)
%     \Lambdam_{k} \nn\\
%     &\times\Big(\Pt^*_k \otimes \Ct_{k}\Big) \Big)^{-1}\Big(\Pt^T_k\otimes \Ct^H_{k}\Big) \Lambdam_k\Big(\Pt_{k}^* \otimes \Ct_k\Big) \nn\\ &\hspace{-0.9cm} \times \Big(\I +\Big(\Pt_{k}^T \otimes \Ct_k^H\Big) \Lambdam_k\Big(\Pt_{k}^* \otimes \Ct_k\Big)\Big)^{-1}\Big(\Pt_{k}^T \otimes \Ct_k^H\Big) \Lambdam_k^{1/2}\Big) \nn\\
%    \st \quad & \Pt_k^H \Pt_k = \I \nn\\
%     &\Ct_k^H \Ct_k = \I .
%\end{align}
%\begin{rem}
%Note that in problem \eqref{eq:prob1}, the secret key rate depends only on the Kronecker product $\Pt^*_k\otimes\Ct_{k}$.

%Define the Kronecker product $\Big(\Pt^*_k\otimes\Ct_{k}\Big)$ as $\U_k$, which is also a tall unitary matrix.
%First, we design the the matrix $\U_k$ that realizes the maximal secret key rate. Based on the optimal $\U_k$, we then
In order to mitigate the interference, the precoding and receiving matrices must satisfy the interference neutralization constraint. Therefore, a novel algorithm will be presented to help design these matrices.
%Generally, the precoding and receiving matrices contain the transmit directions as well as the transmitted power on each direction.
%To reduce the interference, we simply consider equal power allocation of each direction and mainly focus on designing the transmit direction. Therefore the optimization problem can be expressed as
%
%\begin{align}\label{eq:p0}
%    \max_{\Pt_k,\Ct_k} \quad & R_{\text{sum}} = \sum_k I_k \nn\\
%    \st \quad &  \Pt_k^H \Pt_k = \I ,\quad \Ct_k^H \Ct_k = \I,
%\end{align}

\subsection{The IMBA Algorithm} % (fold)
\label{sec:CDR-KGS}

Referring to Proposition 1, as the number of antennas tends to infinity, different elements of the beam domain channel matrix $\Ht_k$ represent the channel gains from different AoDs to different AoAs, which indicates that the channel gains are concentrated in a few beams.
Specifically, suppose that there are $N_P$ paths, each corresponding to different AoAs and AoDs. Then, the BS selects the strongest $N_P$ non-overlapping beams, i.e., the precoding matrix $\Pt_k$ is given by
\begin{align}\label{eq:pt}
    \Pt_k = \begin{bmatrix}
        \e_{\eta_{t,k,1}} & \e_{\eta_{t,k,2}} & \cdots & \e_{\eta_{t,k,N_P}}
    \end{bmatrix}
\end{align}
where $\eta_{t,k,1}$ is the index of the sorted eigenvalue of matrix $\R_{BS,k}$.
Similarly, UT $k$ selects the strongest $N_P$ non-overlapping receiving directions, i.e., the receiving matrix $\Ct_k$ is given by
\begin{align}\label{eq:ct}
    \Ct_k = \begin{bmatrix}
        \e_{\eta_{r,k,1}} & \e_{\eta_{r,k,2}} & \cdots & \e_{\eta_{r,k,N_P}}
    \end{bmatrix}
\end{align}
where $\eta_{r,k,1}$ is the index of the sorted eigenvalue of matrix $\R_{UT,k}$.

{Recalling \eqref{eq:49}, since the BS and UTs all select non-overlapping beams, the designed precoding and receiving matrices can satisfy the constraint of interference neutralization approach.}

\begin{algorithm}[!h]
\caption{IMBA algorithm.}
\label{alg:2}
\begin{algorithmic}[1]
\Require $\R_{BS,k}$ and $\R_{UT,k}$
\Ensure $\P_{k}$ and $\C_k$
\State \textbf{At the BS side:}
\For{$k=1:K$}
\State  Calculate the beam domain channel covariance matrix $\Rt_{BS,k}$ according to
%\eqref{eq:019}
$\Rt_{BS,k}= \Eb \{ (\Ht_{k}^{DL})^H \Ht_{k}^{DL} \}$
.
\State Select the strongest non-overlapping beams $\Pt_k$ according to \eqref{eq:pt} and \eqref{eq:49}.
\State Construct the precoding matrix $\P_k=\A_{BS}\Pt_k$.
\EndFor
\State \textbf{At the UT side:}
\For{$k=1:K$}
\State  Calculate the beam domain channel covariance matrix $\Rt_{UT,k}$ according to
%\eqref{eq:019}
$\Rt_{UT,k}=\Eb \{  \Ht_{k}^{DL} (\Ht_{k}^{DL})^H \}$
.
\State Select the strongest beams $\Ct_k$ according to \eqref{eq:ct}.
\State Construct the receiving matrix $\C_k=\A_{UT}\Ct_k$.
\EndFor
\end{algorithmic}
\end{algorithm}

The number of paths $N_P$ is relatively small, and $M_e$ and $N_e$ can be chosen equal to the number of paths.
Using the precoding and receiving matrices, we can construct $\Pt_k^* \otimes \Ct_k$ to obtain the $N_P^2$ elements in $\Lambdam_k$, which contains the channel information of the $N_P$ paths.
The proposed approach can largely reduce the pilot overhead and work efficiently in massive MIMO channel model and precoding \cite{6717211}.
Based on the analysis above, we propose a novel IMBA algorithm, which is illustrated in Algorithm \ref{alg:2}.

%It is noteworthy that design of the  matrices $\Pt_k$ and $\Ct_k$  depends on the beam domain covariance matrix $\Rt_{BS,k}$ and $\Rt_{UT,k}$}. As $\Rt_{BS,k}$ and $\Rt_{UT,k}$ are the statistical CSI which remains the same on a larger time scale than the instantaneous CSI, designing $\Pt_k$ and $\Ct_k$ for each secret key generation round is not necessary. Also note that an offline design is possible. Moreover, we can choose the corresponding pilots depending on the statistical CSI scenario.

\section{Numerical Results} % (fold)
\label{sec:numerical_results}
In the simulations, we assume that a BS simultaneously communicates with $K=6$ UTs. The BS is equipped with $M=128$ antennas and each UT is equipped with $N_k=4$ antennas. Furthermore, we assume that the BS and UTs employ ULA with $0.5 \lambda$ antenna spacing and the number of channel paths is $N_P=6$ for each channel between the BS and UTs.
According to \eqref{eq:1}, we can generate a channel with randomly distributed AoDs and AoAs.

\begin{figure}[!t]
\centering
{\includegraphics[width=0.4\textwidth]{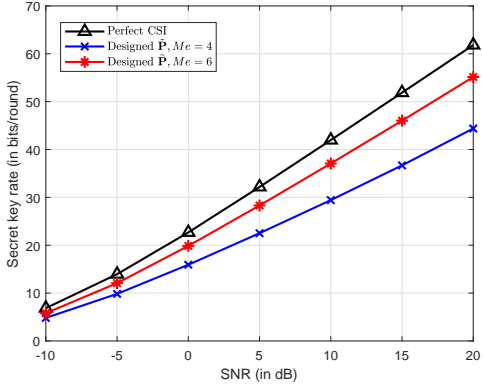}}
\caption{Secret key rate comparison for one UT.}\label{fig:F1}
\end{figure}

First, we evaluate the performance of the beam domain secret key generation scheme in the single user scenario. Fig.~\ref{fig:F1} presents the secret key rate of single user, confirming that the proposed scheme can effectively reduce the dimension of the large channel matrix. The perfect CSI provides the complete channel information and achieves the highest secret key rate. We make a comparison between the secret key rate of the perfect CSI and that of our designed precoding matrices.
Here, we consider $M_e=4$ and $M_e=6$ cases. The numerical results demonstrate that, when $M_e=6$, the secret key rate of designed matrices can approach the perfect case. This indicates that employing the precoding matrix $\Pt$ enables the BS and the UT to obtain the almost perfect channel information, while significantly reducing the pilot overhead and the dimension of channel estimation. When $M_e = 4$, the secret key rate is a little smaller than that of $M_e = 6$, which contains the most channel power with lower overhead.

\begin{figure}[!t]
\centering
{\includegraphics[width=0.4\textwidth]{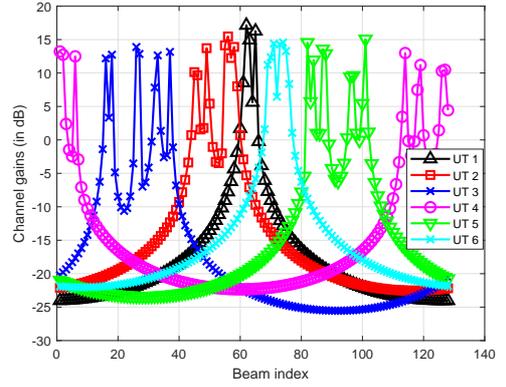}}
\caption{Multi-user channel gains distribution in the beam domain.}
\label{fig:Beam4}
\end{figure}

Next, we consider the multi-user secret key generation and illustrate an example of multi-user channel gains distribution in the beam domain in Fig.~\ref{fig:Beam4}.
The BS employs $128$ antennas to generate $128$ beams with different directions
and the beam index $m$ represents the $m$th beam with direction $\sin(\varphi_m) = 2m/M-1$. When $6$ UTs are distributed in different positions, the channel gains of each UT are concentrated within a few beams, different UTs occupy non-overlapping channel beams. The attenuation between the adjacent UTs is about 20~dB, significantly reducing inter-user interference.
This result indicates that the BS equipped with massive antennas has the potential to achieve multi-user secret key generation.

\begin{figure}[!t]
\centering
{\includegraphics[width=0.4\textwidth]{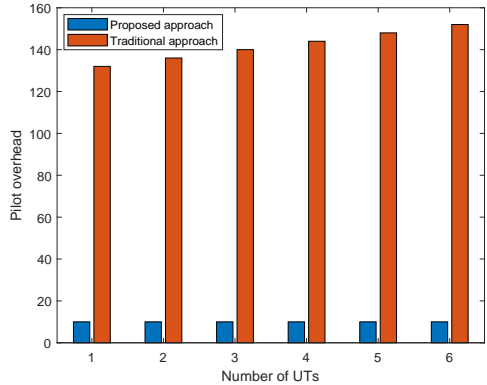}}
\caption{Pilot overhead for traditional and proposed approaches.}\label{fig:F6}
\end{figure}
%
%Fig.~\ref{fig:F4} compares the net secret key rate of overlapping and non-overlapping transmitting beam schemes, when the channel beams are overlapping between different UTs. For the overlapping scheme, the BS allocates the strongest transmitting beams for each UT, some of which may be overlapping with other UTs, while for the non-overlapping scheme, the BS allocates the non-overlapping strongest transmitting beams for each UT.
%For the overlapping transmitting beam scheme, to estimate the channel of overlapping beams for different users, orthogonal pilot is used. Thus, the overhead is a little larger than that of non-overlapping transmitting beam scheme. Here, we do not consider the information leakage of the overlapping beams and only consider the interference between users. We observe that the net secret key rates of non-overlapping schemes are higher than that of overlapping schemes. The non-overlapping scheme with $M_e=6$ achieves the highest rate.
Fig.~\ref{fig:F6} compares the pilot overhead of traditional and proposed approaches multi-user secret key generation.
We observe that due to the large number of antennas at the BS, the traditional overhead $T_{TA}$ is extremely large, meanwhile the overhead also scales with the number of UTs. In contract, the overhead of proposed approach with pilot reuse  $T_{PA}$ remains the same and is significantly lower than that of the traditional approach.
This result demonstrates the desirable performance of the proposed approach in reducing pilot overhead.

\begin{figure}[!h]
\centering
{\includegraphics[width=0.4\textwidth]{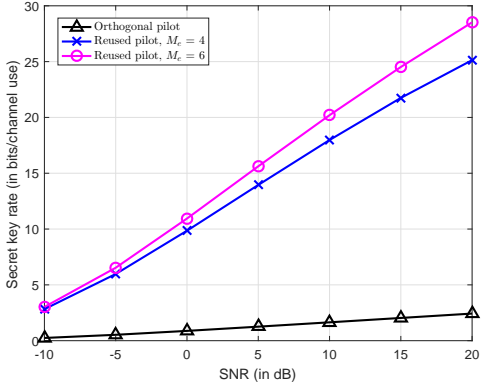}}
\caption{Secret key rate comparison for multiple UTs of orthogonal pilot and reused pilot.}\label{fig:F3}
\end{figure}

 {Since the bottleneck is the pilot overhead in massive MIMO network, we must compare the secret key rate as well as the pilot overhead. }Therefore, we define the unit secret key rate as $R_{\text{unit}} = R_{\text{sum}}/T$,
where $T$ ($T_{TA}$ or $T_{PA}$) is the pilot overhead, scaled with the dimension of the effective channel $M_e$ and $N_e$.
As the number of antennas at each UT is $4$, we set $N_e=N_k=4$.
Fig.~\ref{fig:F3} compares the unit secret key rate of reused pilot with $M_e=4$ and $M_e=6$ with orthogonal pilot scheme. The unit secret key rate in orthogonal pilot schemes suffers serious loss due to its extremely large pilot overhead. The reused pilot scheme with $M_e=6$ achieves the highest rate and the scheme with $M_e=4$ is close to that of $M_e=6$.

\section{Conclusion} % (fold)
\label{sec:conclusion}
This paper provided a design and analysis of the multi-user secret key generation in massive MIMO wireless communications. Exploiting the sparse property of the beam domain channel model, we proposed a channel dimension reduction algorithm named BCP to significantly reduce the dimension of the channel estimation.
Furthermore, we presented a novel algorithm named IMBA which designs the precoding and receiving matrices to support multi-user key generation.
Numerical results demonstrated the performance improvement of our proposed multi-user secret key generation scheme.

\appendices
%\section{Proof of Proposition \ref{prop:1}} % (fold)
%\label{sec:proof_of_proposition_prop:1}
%
%
%From \eqref{eq:6}, the $(n,m)$th element of the beam domain channel $\Ht_{k}^{DL}$
%can be expressed as
%\begin{align}
%    &[\Ht_{k}^{DL}]_{n,m} = \a_{UT,k}(\theta_n)^H \H_{k}^{DL} \a_{Bs}(\varphi_m) \nn\\
%    &= \sum_p \alpha_{k,p} \a_{UT,k}(\theta_n)^H  \a_{UT,k}(\theta_{k,p})  \a_{BS}(\varphi_{k,p})^H   \a_{BS}(\varphi_m).
%\end{align}
%First, we consider the calculation of $\a_{UT,k}(\theta_n)^H  \a_{UT,k}(\theta_{k,p})$.
%As the number of UT antennas tends to infinity,
%there exists $\theta_n$ equal to $\theta_{k,p}$ ($\theta_n = \theta_{k,p}$),
%and
%\begin{align}
%    \a_{UT,k}(\theta_n)^H  \a_{UT,k}(\theta_{k,p}) = 1.
%\end{align}
%When $\theta_n$ is not equal to $\theta_{k,p}$,
%we have \cite{7227112}
%\begin{align}
%    &\lim_{N\to\infty }\a_{UT,k}(\theta_n)^H  \a_{UT,k}(\theta_{k,p})\nn\\& = \lim_{N_k\to\infty }\frac{1}{N_k} \frac{1- e^{-j\frac{2\pi}{\lambda}d N_k (\sin(\theta_{k,p})-\sin(\theta_n))}}{{1- e^{-j\frac{2\pi}{\lambda}d(\sin(\theta_{k,p})-\sin(\theta_n))}}} =0.
%\end{align}
%Similarly, as the number of BS antennas grows,
%we have
%\begin{align}
%    \lim_{M\to\infty} \a_{BS}(\varphi_{k,p})^H   \a_{BS}(\varphi_m) = \delta(\varphi_{k,p}- \varphi_{m}).
%\end{align}
%Thus, the $(n,m)$th element of $\Ht_{k,p}$ can be expressed as
%\begin{align}
%    \lim_{N,M\to\infty} [\Ht_{k}^{DL}]_{n,m} -\sum_p \alpha_{k,p} \delta(\theta_k,p-\theta_n)\delta(\varphi_{k,p}- \varphi_{m}) = 0.
%\end{align}
%This completes the proof.\qed

\section{Proof of Theorem \ref{thm:1}} % (fold)
\label{sec:proof_of_theorem_thm:1}

We assume zero-mean complex Gaussian random vector for each channel observation $\z_k^{DL}$ or $\z_k^{UL}$.
% \begin{align}\label{eq:I1}
%     I (\z_k^{DL};\z_k^{UL} | \z_i^{DL}  )
%     =& H(\z_k^{DL}, \z_i^{DL}) + H(\z_k^{UL} , \z_i^{DL} )\nn\\
%     &\quad- H(\z_k^{DL}, \z_k^{UL} , \z_i^{DL}) - H(\z_i^{DL}) \nn\\
%     =& \log \frac{\det( \Rc_{\z_k^{DL} \z_i^{DL}} \Rc_{\z_k^{UL} \z_i^{DL}} )}{\det( \Rc_{\z_k^{DL} \z_k^{UL} \z_i^{DL}}) \det(\Rc_{ \z_i^{DL}} )}.
% \end{align}
When the channel observations of different UTs are uncorrelated, we have \cite{5483148}
% the conditional mutual information \eqref{eq:I1} can be simplified as
\begin{align}
    I_k = I (\z_k^{DL};\z_k^{UL}  )
    =  \log \frac{\det( \Rc_{\z_k^{DL} } \Rc_{\z_k^{UL}} )}{\det( \Rc_{\z_k^{DL} \z_k^{UL} } )},
\end{align}
which only depends on the correlation of uplink and downlink channels.

Let $\R_k = \Eb\{\tvec(\H_k) \tvec(\H_k)^H\}$ be the full correlation of the channel matrix. We can calculate $\Rc_{ \z_k^{DL}} $ as
\begin{align}
    &\Rc_{ \z_k^{DL}} = \sum_{k'}((\P_{k'})^T \otimes \C_k^H) \R_k \sum_{k'}((\P_{k'})^T \otimes \C_k^H)^H\nn\\
    & + (\I_{T_D} \otimes \C_k^H \C_k)  .
\end{align}
Note that $\R_k$ can be decomposed as $\R_k = (\A_{BS}^* \otimes \A_{UT}) \Lambdam_{k} (\A_{BS}^* \otimes \A_{UT})^H$.
Let $\Pt_k = \A_{BS}^H\P_{k}$, $\Ct_k = \A_{UT,k}^H\C_k$, $\V_k = \Lambdam_k^{1/2}\Big(\sum_{k'}(\Pt_{k'})^T \otimes \Ct_k^H\Big)^H$ and $\V_{kk'} = \Lambdam_{k'}^{1/2} \Big(\Pt^T_k\otimes \Ct^H_{k'}\Big)^H$.
The covariance matrix $\Rc_{ \z_k^{DL}}$ can be rewritten as
\begin{align}
    \Rc_{ \z_k^{DL}} =\V_k^H \V_k+ (\I_{T_D} \otimes \C_k^H \C_k).
\end{align}
Similarly, we can calculate $\R_{\z_k^{UL}}$ and $\R_{\z_k^{DL} \z_k^{UL}}$ as
\begin{align}
    \Rc_{ \z_k^{UL}} =&\sum_{k'} \V_{kk'}^H \V_{kk'} + (\P^T_k\P^*_k \otimes \I_{T_U})\nn\\
    \R_{\z_k^{DL} \z_k^{UL}} =& \V_k^H \V_{kk}.
\end{align}

The covariance matrix $\Rc_{\z_k^{DL}\z^{UL}_k}$ can be decomposed as
\begin{align}
    \Rc_{\z_k^{DL}\z^{UL}_k} = \begin{bmatrix}
        \Rc_{\z_k^{DL}} & \R_{\z_k^{DL} \z_k^{UL}}  \\
        \R_{\z_k^{UL} \z_k^{DL}}  &  \Rc_{\z_k^{UL}}.
    \end{bmatrix}
\end{align}
From the determinant of the block matrix, we have
\begin{multline}
    \det(\Rc_{\z_k^{DL}\z^{UL}_k}) = \det(\Rc_{\z_k^{DL}})\\
    \times\det\Big(\Rc_{\z_k^{UL}} - \R_{\z_k^{UL} \z_k^{DL}}\Rc_{\z_k^{DL}}^{-1}\R_{\z_k^{DL} \z_k^{UL}} \Big).
\end{multline}
Then, the secret key rate can be expressed as
%\begin{align}\label{eq:57}
 %   I_k
%    = -\log \det \left( \I - \Rc_{\z_k^{UL}}^{-1}\R_{\z_k^{UL} \z_k^{DL}}\Rc_{\z_k^{DL}}^{-1}%\R_{\z_k^{DL} \z_k^{UL}} \right),
%\end{align}
%which is given by
\begin{align}
  &I_k = \!- \log\! \det\! \Bigg(\!\I - \V_{kk}\!  \left(\! \sum_{k'} \V_{kk'}^H \V_{kk'}  \!+\! \Big(\P^T_k\P^*_k \otimes \I_{T_U}\!\Big)\!\!\right)^{-1}\nn\\
    &\quad \times \V_{kk}^H \V_k \left(  \V_k^H\V_k + \I_{T_D} \otimes \C_k^H \C_k \right)^{-1}\V_k^H \Bigg).
\end{align}
This completes the proof. \qed

\section*{Acknowledgment}
This research was supported by the NSFC (61801115), the Zhishan Youth Scholar Program Of SEU, the Research Fund of National Mobile Communications Research Laboratory，Southeast University (No.2019B01) and the Fundamental Research Funds for the Central Universities (3204009415, 3209019405). The work of E. Jorswieck is partly funded by the German Research Foundation (DFG) under project JO 801/21-1. The work of B. Xiao is partly funded by the NSFC (61772446).
\bibliographystyle{IEEEtran}
\bibliography{IEEEabrv,mybibfile}

% that's all folks
\end{document}